\documentclass[conference]{IEEEtran}

\usepackage[export]{adjustbox}
\usepackage{algorithm}
\usepackage{algorithmic}
\usepackage{amsmath,amssymb,amsfonts}
\usepackage{cite}
\usepackage{float}
\usepackage{graphicx}
\usepackage{listings}
\usepackage{textcomp}
\usepackage[table,xcdraw]{xcolor}
\usepackage[normalem]{ulem}
\usepackage{url}
\usepackage{tcolorbox}
\useunder{\uline}{\ul}{}

\newtheorem{mydef}{Definition}

\def\BibTeX{{\rm B\kern-.05em{\sc i\kern-.025em b}\kern-.08em
    T\kern-.1667em\lower.7ex\hbox{E}\kern-.125emX}}

\begin{document}

\title{Verifying Security Vulnerabilities in Large Software Systems using Multi-Core \textit{k}-Induction}

\author{\IEEEauthorblockN{Thales Silva, Carmina Porto, Erickson Alves}
\IEEEauthorblockA{\textit{Federal University of Amazonas}\\
Manaus, Brazil \\
\{thales.tas, cdessana, erickson.higor\}@gmail.com}
\and
\IEEEauthorblockN{Lucas C. Cordeiro}
\IEEEauthorblockA{
\textit{University of Manchester}\\
Manchester, UK\\
lucas.cordeiro@manchester.ac.uk}
\and
\IEEEauthorblockN{Herbert O. Rocha}
\IEEEauthorblockA{
\textit{Federal University of Roraima}\\
Roraima, Brazil\\
herbert.rocha@ufrr.br}
}

\maketitle

\begin{abstract} 
Computer-based systems have been used to solve several domain problems, such as industrial, military, education, and wearable. Those systems need high-quality software to guarantee security and safety. We advocate that Bounded Model Checking (BMC) techniques can detect security vulnerabilities in the early stages of development processes. However, this technique struggles to scale up and verify large software commonly found on computer-based systems. Here, we develop and evaluate a pragmatic approach to verify large software systems using a state-of-the-art bounded model checker. In particular, we pre-process the input source-code files and then guide the model checker to explore the code systematically. We also present a multi-core implementation of the \textit{k}-induction proof algorithm to verify and falsify large software systems iteratively. Our experimental results using the Efficient SMT-based Model Checker (ESBMC) show that our approach can guide ESBMC to efficiently verify large software systems. We evaluate our approach using the PuTTY application to verify $\mathbf{136}$ files and $\mathbf{2803}$ functions in less than $\mathbf{86}$ minutes, and the SlimGuard allocator, where we have found real security vulnerabilities confirmed by the developers. We conclude that our approach can successfully guide a bounded model checker to verify large software systems systematically.
\end{abstract}


\section{Introduction} \label{sec:introduction}

In our society, problems of several domains (\textit{e.g.}, industrial,
military, education, and wearable) have been solved by applying computer-based
systems, which usually require high-quality software to satisfy a set of safety
and security requirements by such systems~\cite{CordeiroFB20}. Furthermore, critical embedded
systems, such as those in the industrial domains, impose several security properties
(\textit{i.e.}, confidentiality, integrity, and availability). According to
systems' requirements, these properties must be met and validated and need to be
performed at the early stages of development processes; if not, failures may
lead to catastrophic situations, \textit{e.g.}, loss of money and life. As an outcome,
software testing and verification techniques are essential tools for developing
systems with high dependability and reliability requirements, thereby ensuring
both user requirements and system behavior are fundamental.

In critical domains, the software must be carefully tested so that violations are not observed during the execution in real scenarios~\cite{Myers2004}. For instance, in the C programming language~\cite{Kernighan2006}, which is widely used to develop critical software, \textit{e.g.}, operating systems and drivers, execution of unsafe code might lead to an undefined behavior~\cite{related-2}, which are common causes of memory errors, \textit{e.g.}, buffer overflows and double-free violations. Furthermore, such types of errors are one of the main threats to software security~\cite{Veen2012}, since they can be exploited by attackers to execute malicious code. When it comes to open-source software, such vulnerabilities increase since attackers can quickly read the code and find failures.

BMC techniques can falsify properties up to a given depth $k$; however, they cannot prove system correctness unless an upper bound of $k$ is known, \textit{i.e.}, a given bound that unwinds all loops and recursive functions to their maximum possible depths. BMC techniques limit the visited regions of data structures and the number of related loop iterations. Therefore, BMC limits the state-space that needs to be explored during verification in such a way that real errors in applications can be found~\cite{Clarke2004, Merz2012, Gadelha2019, Ivancic2005}. Nevertheless, BMC tools are susceptible to exhaustion of time and/or memory limits when verifying programs with loop bounds that are too large.

Several techniques have been proposed to address software vulnerabilities in C programs. For instance, fuzzing~\cite{Bohme2017,Godefroid20} techniques, including black-, grey- or white-box,  have been widely used to exploit inputs for programs that lead to unexpected behaviors. Another example of approaches that aim at finding violations in software is the use of static analysis tools that check C programs~\cite{CadarDE08,related-1, Clarke2004, Gadelha2019} concerning safety properties. Semiformal verification approaches have also been proposed that combine static and dynamic verification~\cite{CordeiroFCM09}. However, such techniques struggle to scale up and verify large software commonly found in open-source applications. Recently, Cook et al.~\cite{Cook2020} describe the use of bounded model checkers to triage the severity of security bugs at the cloud service provider Amazon Web Services (AWS).  However, several abstractions performed in the verification approach might cause the model checker to miss traces and not automatically falsify spurious traces.

We propose a pragmatic approach to verify large software systems using a state-of-the-art bounded model checker. In particular, we pre-process the input source-code files and then guide the model checker to explore specific locations in the state-space, \textit{e.g.}, functions, so that the usage of the computational resources is reduced and it can find real bugs in large software systems. Once a counterexample is found, we use existing approaches to check whether such violation is reproducible in the program or not~\cite{understanding,0001S20}. Our main contribution is to propose a novel approach for detecting and exploiting security vulnerabilities in large software systems. We leverage the benefits of using BMC techniques to detect security vulnerabilities hidden deep in the software state-space and existing counterexample validators to eliminate spurious bugs. In particular, we make the following significant contributions:

\begin{itemize}

\item Provide a novel verification approach that combines code analysis and BMC
	techniques to detect software vulnerabilities in open-source applications and
		assess findings in a detailed, user-friendly report.

\item Present a multi-core implementation of the $k$-induction proof algorithm. It exploits incremental BMC to iteratively verify and falsify programs for each unwind bound indefinitely (\textit{i.e.}, until it exhausts the time or memory limits), thereby enabling BMC techniques to verify programs up to three times faster than sequential implementations.

\item Experimental results show that our pragmatic verification approach can
	find security vulnerabilities, \textit{e.g.}, memory leaks, and pointer-safety
		violations in open-source C programs. Furthermore, some of the
		vulnerabilities reported by our approach have been confirmed by the
		developers.
\end{itemize}

\section{Background}
\label{sec:background}

\subsection{Security Vulnerabilities}

As a formal definition, a vulnerability in a program is a property that violates
the Confidentiality, Integrity, and Availability (CIA) triad, thereby allowing
sensitive data to be leaked or even malicious code to be written as part of the
program~\cite{LundgrenM19}. The CWE Community identifies the most common
vulnerabilities regarding the C/C++ programming language~\cite{cwe}. Here we
describe four of them, which we consider in this study.   

\begin{mydef}
\label{BufferOverflow}
(\textit{Buffer Overflow}) A buffer overflow occurs when data are written to a
	buffer also corrupts data values in memory addresses adjacent to the
	destination buffer due to insufficient bounds checking. This vulnerability can
	occur when copying data from one buffer to another without first checking
	whether it fits within the destination buffer~\cite{cwe}.
\end{mydef}

\begin{mydef}
\label{ImproperBufferAccess}
(\textit{Improper Buffer Access}) The software uses a sequential operation to
	read or write a buffer, but it uses an incorrect length value, which causes it
	to access memory outside of the buffer's bounds. When the length value exceeds
	the size of the destination, a buffer overflow could occur~\cite{cwe}.
\end{mydef}

\begin{mydef}
\label{NullPointerDerefence}
(\textit{Null Pointer Derefence}) A NULL pointer dereference occurs when the
	application dereferences a pointer that it expects to be valid, but is NULL,
	typically causing a crash or exit. NULL pointer dereference issues can occur
	through many flaws, including race conditions and simple programming
	omissions~\cite{cwe}.
\end{mydef}

\begin{mydef}
\label{DoubleFree}
(\textit{Double Free}) When a program calls \textit{free()} twice with the same
	argument, the program's memory management data structures become corrupted.
	This corruption can cause the program to crash or, in some circumstances,
	cause two later calls to \textit{malloc()} to return the same pointer. Suppose
	\textit{malloc()} returns the same value twice, the program later gives the
	attacker control over the data written into this doubly-allocated memory. In
	that case, the program becomes vulnerable to a buffer overflow
	attack~\cite{cwe}. 
\end{mydef}

\subsection{ESBMC}
\label{sec:esbmc}

Bounded model checking (BMC) techniques, both based on Boolean Satisfiability
(SAT)~\cite{Biere2009} or Satisfiability Modulo Theories (SMT)~\cite{Moura2009},
have been successfully applied to verify single and multi-threaded programs and
find subtle bugs in real programs~\cite{Clarke2004, Merz2012, Gadelha2019,
Byer2015, Carter2016}. The general idea behind BMC is to check the negation of a
given property at a given depth, \textit{i.e.}, given a transition system $M$, a
property $\phi$, and a limit of iterations $k$, BMC unfolds a given system $k$
times and converts it into a verification condition (VC) $\psi$, such that
$\psi$ is \textit{satisfiable} \textit{iff} $\phi$ has a counterexample of
depth less than or equal to $k$.

ESBMC is a mature bounded model checker that supports the verification of
single- and multi-threaded C/C++ programs. It can automatically check both
predefined safety properties (\textit{e.g.}, memory leaks, pointer safety, and
overflows) and user-defined program assertions. It has been awarded several
prizes at the International Competition on Software Verification
(SV-Comp)~\cite{Dirk2020} and has been applied to several domains
(\textit{e.g.}, digital controllers~\cite{Cavalcante2020}, photovoltaic
systems~\cite{Trindade2019}). Nonetheless, as a bounded model
checker, it is susceptible to resource exhaustion problems, and scaling up BMC
tools to handle large software is an open problem.

\section{Speeding Up BMC by Using Parallelism and Incremental Verification}
\label{sec:speeding-up-bmc}

\subsection{Incremental Verification}
\label{sec:incremental-verification}

Our incremental verification algorithm works as follows~\cite{Gadelha2019}. Let a given C program $P$ be
modeled as a finite transition system $M$, which is defined as follows:

\begin{itemize}

\item[--] $I(s_n)$ and $T(s_n, s_{n+1})$ as the formulae over program's state
	variable set $s_i$ constraining the initial states and transition relations of
		$M$;

\item[--] $\phi(s)$ as the formula encoding states satisfying a required safety
	property, which aims to verify language-specific properties and user-defined
		properties;

\item[--] $\psi(s)$ as the formula encoding states satisfying the completeness
	threshold, i.e., states corresponding to the termination. $\psi(s)$
		contains unwindings no more profound than the maximum number of
		loop-iterations in $P$.

\end{itemize}

Note that, in our notation, termination and error are mutually exclusive:
$\phi(s) \wedge \psi(s)$ is by construction unsatisfiable; $s$ is a deadlock
state if $T(s, s') \vee \phi(s)$ is unsatisfiable.

In each step $k$ of the incremental verification algorithm, two propositions are
checked: the base case $B(k)$ and the forward condition $F(k)$. $B(k)$
represents the standard BMC and it is satisfiable \textit{iff} $P$ has a
counterexample of length \textit{k} or less:
\begin{equation}\label{eq:bk}
 B(k) \Leftrightarrow I(s_1) \wedge \left( \bigwedge^{k-1}_{i=1} T(s_i, s_{i+1}) \right) \wedge
\left( \bigvee^{k}_{i=1} \neg \phi(s_i) \right) .
\end{equation}

The forward condition checks for termination, i.e., whether the completeness
threshold $\psi(s)$ must hold for the current $k$.  If $F(k)$ is unsatisfiable,
$P$ has terminated:
\begin{equation}\label{eq:fk}
 F(k) \Leftrightarrow I(s_1) \wedge \left( \bigwedge^{k-1}_{i=1} T(s_i, s_{i+1}) \right) \wedge \neg
\psi(s_k).
\end{equation}

No safety property $\phi(s)$ is checked in $F(k)$ as they were checked for the
current \textit{k} in the base case. Finally, the inductive condition $S(k)$ is
unsatisfiable if, whenever $\phi(s)$ holds for $k$ unwindings, it also holds for
the next unwinding of $P$:
\begin{equation}\label{eq:sk}
 S(k) \Leftrightarrow \exists n \in \mathbb{N}^{+} . \bigwedge^{n+k-1}_{i=n} (\phi(s_{i}) \wedge T'(s_i, s_{i+1})) \wedge \neg \phi(s_{n+k}).
\end{equation}
Here $T'(s_i, s_{i+1})$ is the transition relation after havocking the loop
variables~\cite{Gadelha2017}.

Through $B(k)$, $F(k)$, $S(k)$, and $\pi(k) \Leftrightarrow B(k) \vee [ F(k)
\wedge S(k) ] $, the incremental verification algorithm
$\mathrm{bmc}_{\mathrm{inc}}$ to falsify or verify programs at a given $k$ is:
\begin{equation}\label{eq:verk}
 \mathrm{bmc}_{\mathrm{inc}}(P, k) =
 \begin{cases}
 P \text{ is unsafe}, & \text{if}\ B(k)\ \text{is SAT},\\
 P \text{ is safe}, & \text{if}\ \pi(k) \ \text{is UNSAT},\\
 \mathrm{bmc}_{\mathrm{inc}}(P, k+1), & \text{otherwise}.
 \end{cases}
\end{equation}

\subsection{Proof by \textit{k}-Induction in Multi-core Architectures}

Algorithm~\ref{alg:kind-algorithm} shows an overview of the proposed $k$-induction algorithm. The input of the algorithm is a C program $P$ along with a property $\phi$. The algorithm returns \textit{true} (in case it could not find a path that violates $\phi$), \textit{false} (in case it could find such a path), and \textit{unknown} (in case it does not succeed in computing the previous answers).

In the base case, the $k$-induction algorithm aims at finding a counterexample up to a given maximum number of iterations $k$. In the forward condition, global loop correctness w.r.t. $\phi$ is shown for the case, which the loop iterates at most $k$ times. In the inductive step, the algorithms check that, whenever $\phi$ is valid in $k$ iterations, it must also be valid for the next iterations. The algorithm runs up to a maximum number of iterations, and $k$ is increased if it cannot falsify $\phi$ during the base case.

\begin{algorithm}[htb]
\caption{Proposed $k$-induction algorithm.}
\label{alg:kind-algorithm}
\centering
\begin{algorithmic}
\REQUIRE {program $P$ and a safety property $\phi$}
\ENSURE {true, false, or unknown}
\STATE {$k \gets 1$}
\WHILE {$k \leq max\_iterations$}
  \IF {$base\_case(P, \phi, k)$}
    \STATE {$counterexample$ $s[0..k]$}
    \RETURN {$false$}
  \ELSE
    \IF {$forward\_condition(P, \phi, k)$}
      \RETURN {$true$}
    \ELSE
      \IF {$inductive\_step(P, \phi, k)$}
        \RETURN {$true$}
      \ENDIF
    \ENDIF
  \ENDIF
\ENDWHILE
\RETURN {$unknown$}
\end{algorithmic}
\end{algorithm}

However, even if we use such an approach to find extensive software violations,
BMC is susceptible to large verification times and even the exhaustion of
time/memory limits. This limitation has motivated this work to develop a
different approach to speed up the BMC procedure and find violations in less
time. Therefore, we have extended previous work~\cite{Gadelha2017}, which
implements Algorithm~\ref{alg:kind-algorithm} by using four different processes:
the primary process $PP$, which is responsible for spawning the other
subprocesses and reasoning about the obtained results from each; and the base
case $BCP$, forward condition $FCP$, inductive step $ISP$ processes, which are
responsible for verifying the program w.r.t. the base case, forward condition,
and inductive step, respectively.

First, the primary process spawns the other three subprocesses, instantiates a two-way pipe for each subprocess, and the verification procedure starts. In each step $k$, the verification occurs as follows: $BCP$ performs the base case procedure as described in Algorithm~\ref{alg:kind-algorithm} at the maximum bound $k$ and if a counterexample has been found, it sends such result to $PP$ via its pipe. $BCP$ performs the forward condition procedure as described in Algorithm~\ref{alg:kind-algorithm} at the maximum bound $k + 1$ and if the program's correctness has been proved, it sends such result to $PP$ via its pipe. Lastly, $ICP$ performs the inductive step procedure as described in Algorithm~\ref{alg:kind-algorithm} at the maximum bound $k + 1$ and if the correctness of the property has been proved, it sends such result to $PP$ via its pipe. Whenever $PP$ receives a result from any subprocess, it kills the others and reports the user's verification result. The pseudocodes for $PP$ and $BCP$ are described in Algorithms~\ref{alg:pp-algorithm} and~\ref{alg:bcp-algorithm}, respectively. We chose to omit the pseudocodes for $FCP$ and $ISP$, since they differ from $BCP$ only when setting the initial parameters and they aim at proving the correctness of the program, as described in Algorithm~\ref{alg:kind-algorithm}. According to our experiments, such an approach might reduce the verification time by a factor of $3$.

\begin{algorithm}[htb]
\caption{$PP$ algorithm.}
\label{alg:pp-algorithm}
\centering
\begin{algorithmic}
\REQUIRE {program $P$ and a safety property $\phi$}
\ENSURE {true, false, or unknown}
\STATE {spawn $BCP$, $FCP$, $ICP$}
\STATE {wait subprocesses execution to finish}
\IF {$BCP$ found a solution}
  \RETURN {$false$}
\ELSE
  \IF {$FCP$ found a solution}
    \RETURN {$true$}
  \ELSE
    \IF {$ISP$ found a solution}
      \RETURN {$true$}
    \ENDIF
  \ENDIF
\ENDIF
\RETURN {$unknown$}
\end{algorithmic}
\end{algorithm}

\begin{algorithm}[htb]
\caption{$BCP$ algorithm.}
\label{alg:bcp-algorithm}
\centering
\begin{algorithmic}
\REQUIRE {program $P$, a safety property $\phi$, and an upper bound $K$}
\ENSURE {bound $k$ which violates $\phi$}
\STATE {$set\_base\_case\_parameters()$}
\WHILE {$k \leq K$}
  \IF {$incremental\_bmc(P, \phi, k)$}
    \RETURN {$k$}
  \ENDIF
\ENDWHILE
\RETURN {$0$}
\end{algorithmic}
\end{algorithm}

In this work, we apply the incremental verification approach described in Sec.~\ref{sec:incremental-verification} on top of the multi-core $k$-induction algorithm to find violations in less time and enable BMC to reason about large software.

\section{Verifying Security Vulnerabilities in Open-Source Applications}
\label{sec:method}

Here, we describe our proposed method to verify open-source applications that consist of multiple files using a software model checker in a completed automated manner. Large C or C++ programs are often divided into multiple files. We can refer to open-source programs such as PuTTY (175 files) \cite{putty}, OpenSSH (276 files) \cite{openssh}, and OpenSSL (1239) \cite{openssl}. The number of files that compose an application can increase considerably depending on the application's complexity. 

Dividing the applications into several files allows keeping each file short enough to be conveniently edited and allows some of the code to be shared with other programs. However, it also makes the verification task more complicated and hard to manage. Each file must be verified and specific dependencies must be included to check applications with several files. Some of these files do not contain the \textit{main} function, which software model checkers usually use to verify that.

\begin{figure*}[h]
	\centering
	\includegraphics[width=12cm]{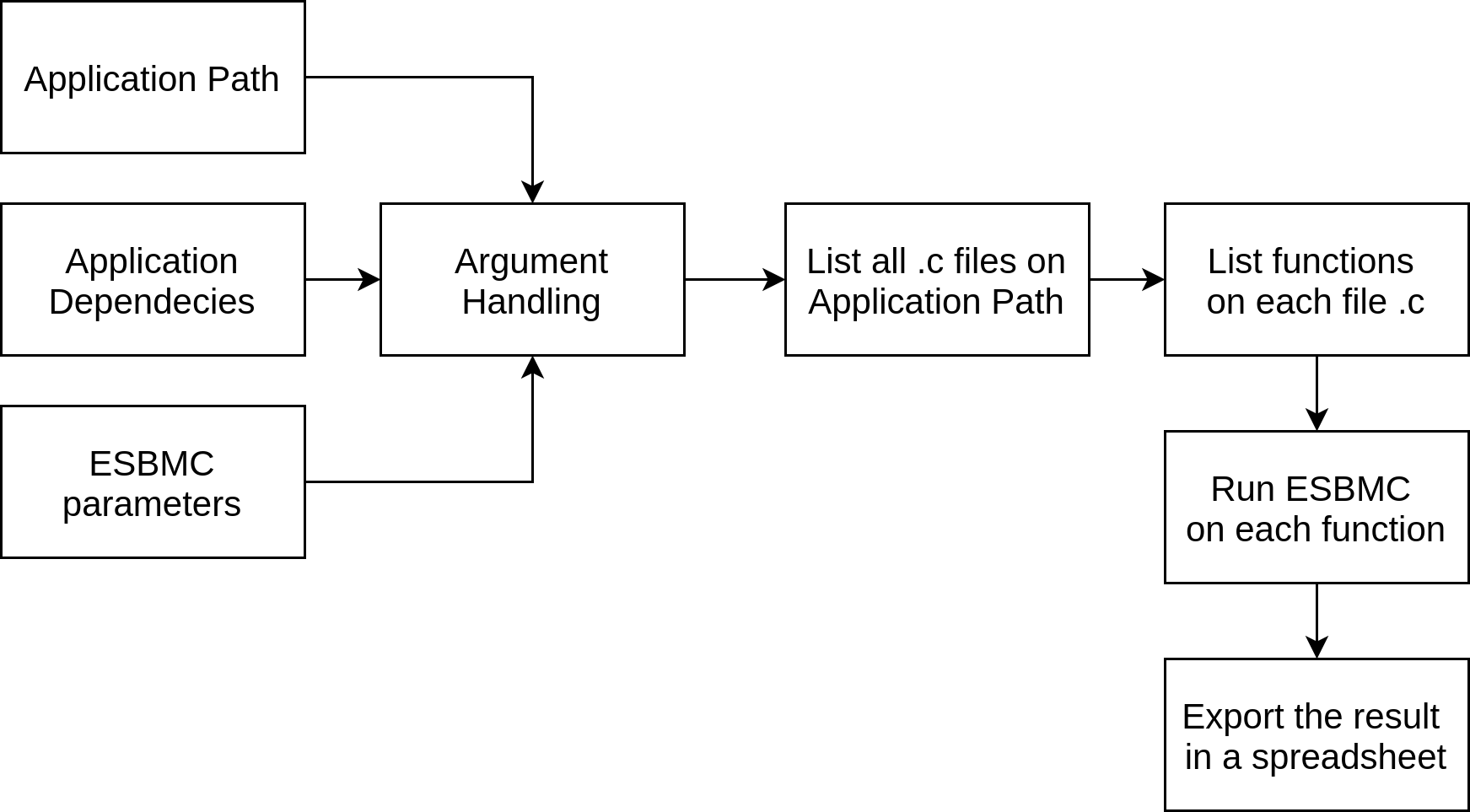}
	\caption{Process steps of the proposed tool - \textit{esbmc-wr}. }
	\label{script-new}
\end{figure*}

So this work presents an automated tool external to a model checker to verify a C program's entire source code. It is called \textit{esbmc-wr}. As described in Algorithm~\ref{alg:script_algorithm}, given a C program source code $P$, and its directory $D$, we first list all $.c$ files in $D$. Each file is analyzed, as we look for all declared functions $F$ on it. Then, we run each $F$ in a model checker to verify different types of violations, such as pointer safety, arithmetic overflow, division by zero, and out-of-bounds arrays. The type of violations is passed to our script as an argument $A$. After the verification process is concluded, we produce a spreadsheet $V$ with the model checker logs' outcomes. Figure~\ref{script-new} illustrates each process step of the script.
\begin{algorithm}[htb]
\caption{Proposed script algorithm.}
\label{alg:script_algorithm}
\centering
\begin{algorithmic}
\REQUIRE {program $P$, its directory D and a set of arguments $A$}
\ENSURE {Verification Outcome V}
	\STATE {$args \gets list\_arguments(A)$}
	\STATE {$files \gets list\_files(P, D)$}
\STATE {$k \gets 1$}
\WHILE {$k \leq files$}
	\STATE {$functions \gets list\_function(files[k])$}
	\STATE {$log \gets ESBMC\_Check(files[k], functions[k], args) $}
	\STATE {$k \gets k + 1$}
\ENDWHILE
\STATE {$V \gets spreedsheat(log)$}
\RETURN {$V$}
\end{algorithmic}
\end{algorithm}

\subsection{Using ESBMC to verify source code files}

In this work, we use the \textit{k}-induction proof rule of ESBMC in multi-core architecture to verify each function listed on the source code files. ESBMC converts C/C++ programs into GOTO programs~\cite{HandlingLoop}. This conversion replaces all control structures by (conditional) jumps, which simplifies the program representation. This GOTO representation's symbolic execution converts the program into a Static Single Assignment (SSA) form. It unrolls loops and recursive functions on-the-fly, thereby generating unwinding assertions that fail if the given bound is too small. The next step is the symbolic execution, when the GOTO program is executed (unrolling loops up to the bound $k$) and converted to SSA form. 

ESBMC performs various optimizations in the SSA level, including constant propagation (to simplify expressions further) and instruction slicing (to remove unnecessary instruction); the simplification process is an important step and speeds up the program verification for some particular applications. We have two slicing strategies. First, we remove all instructions after the last assert in the SSA set. After that, we collect all the symbols (and their dependent symbols) in assertions and remove instructions that do not depend on them. Both strategies ensure that unnecessary instructions are ignored in the next step, the SMT encoding.

The SSA expressions are then encoded using an SMT solver. If the SMT formula is shown to be satisfiable, a counterexample is presented describing the error found. If the formula is found to be unsatisfiable, there are no errors up to the unwinding bound $k$, and this result is presented \cite{soton416918}.

\subsection{Setup Configuration}
\label{sec:SetupConfiguration}

In order to execute our script, we must initiate it on the application path. Users can set the parameters providing this as an argument to the script; most of them are forwarded to the model checker. Each parameter is described as follows. 
\textbf{-m} enables memory leak checks;
\textbf{-u} sets the number of loop unwinding, e.g., -u 50;
\textbf{-nu} disables unwinding assertions;
\textbf{-ib} enables incremental solving;
\textbf{-p} disables pointer check;
\textbf{-o} enables overflow checks;
\textbf{-t} sets the timeout;
\textbf{-i} some files have specific dependencies; our script provides this dependencies to ESBMC through a file. This option is a path to the file that describes the libraries dependencies, e.g., -i dep.txt;
\textbf{-k} enables multi-core \textit{k}-induction;
\textbf{-w} exports a graphml file to perform violation witness checking;
\textbf{-f} enables verification function-by-function;
\textbf{-h} shows the script available options.

The \textit{esbmc-wr} script was developed to work with ESBMC, but it can be modified to work with another software verifier. It is necessary to change the arguments parse to fit the new model checker parameters.

\subsection{File Listing}

We list all source code files available on the directory, where the script was initialized by merely using the function shown in Figure~\ref{list-files}.
\begin{figure}[ht]
\centering
\begin{minipage}{0.45\textwidth}
\begin{lstlisting}[stepnumber=1,
numbers=left,
firstnumber=1,
numberstyle=\tiny,
extendedchars=true,
breaklines=true,
frame=single,
basicstyle=\footnotesize,
stringstyle=\ttfamily,
showstringspaces=false,
captionpos=b,
breakautoindent=truem
language=python,
numbersep=5pt,
tabsize=2,
morekeywords={assert,assume,int,void,NULL,return,for,malloc,include,if,else,free,sizeof,define,struct,while,typedef,goto,short,char,pthread\_create,pthread\_t,pthread\_join,double}]
    def list_c_files():
	    return(glob.glob("*.c"))

\end{lstlisting}
\end{minipage}
\caption{An example of how to use the python script \textit{esbmc-wr}.}
\label{list-files}
\end{figure}
The output of this function is a list with all ``.c'' files found in a directory. Figure~\ref{list-out} illustrates an example of the PuTTY function listing \cite{putty}. 

\begin{figure}[ht]
\centering
\begin{minipage}{0.45\textwidth}
\begin{lstlisting}[stepnumber=1,
numbers=left,
firstnumber=1,
numberstyle=\tiny,
extendedchars=true,
breaklines=true,
frame=single,
basicstyle=\footnotesize,
stringstyle=\ttfamily,
showstringspaces=false,
captionpos=b,
breakautoindent=truem
language=python,
numbersep=5pt,
tabsize=2,
morekeywords={assert,assume,int,void,NULL,return,for,malloc,include,if,else,free,sizeof,define,struct,while,typedef,goto,short,char,pthread\_create,pthread\_t,pthread\_join,double}]

['sshbcrypt.c', 'pinger.c', 'version.c', 'sshccp.c', 'sshshare.c', 'callback.c', 'sshdssg.c', 'ssh1censor.c', 'sshmac.c', 'be_misc.c', ... , 'ssh2bpp.c']

\end{lstlisting}
\end{minipage}
	\caption{Output of the function \textit{list\_c\_files()} .c file listed in a directory.}
\label{list-out}
\end{figure}

\subsection{Function listing}

Our script lists all functions on a source code file, as illustrated in Figure~\ref{list-file-functions}. This functionality guarantees that every function will be verified, including the \textit{main} function.

\begin{figure}[ht]
\centering
\begin{minipage}{0.45\textwidth}
\begin{lstlisting}[stepnumber=1,
numbers=left,
firstnumber=1,
numberstyle=\tiny,
extendedchars=true,
breaklines=true,
frame=single,
basicstyle=\footnotesize,
stringstyle=\ttfamily,
showstringspaces=false,
captionpos=b,
breakautoindent=truem
language=python,
numbersep=5pt,
tabsize=2,
morekeywords={assert,assume,int,void,NULL,return,for,malloc,include,if,else,free,sizeof,define,struct,while,typedef,goto,short,char,pthread\_create,pthread\_t,pthread\_join,double}]
    def list_functions(c_file):
	    process = subprocess.Popen([CTAGS,CTAGS_TAB,CTAGS_FUNC,c_file],
		    stdout=subprocess.PIPE,
			stderr=subprocess.PIPE,
			text=True)

			(stdout,stderr) = process.communicate()

			func_list = row_2_list(stdout) 
			func_list = find_main(func_list)

			return(func_list)

\end{lstlisting}
\end{minipage}
\caption{\textit{esbmc-wr} function listing.}
\label{list-file-functions}
\end{figure}

An example of the PuTTY's \cite{putty} \textit{ssgbcrypt.c} file function listing is illustrated in Figure \ref{list-func-out}. 

\begin{figure}[ht]
\centering
\begin{minipage}{0.45\textwidth}
\begin{lstlisting}[stepnumber=1,
numbers=left,
firstnumber=1,
numberstyle=\tiny,
extendedchars=true,
breaklines=true,
frame=single,
basicstyle=\footnotesize,
stringstyle=\ttfamily,
showstringspaces=false,
captionpos=b,
breakautoindent=truem
language=python,
numbersep=5pt,
tabsize=2,
morekeywords={assert,assume,int,void,NULL,return,for,malloc,include,if,else,free,sizeof,define,struct,while,typedef,goto,short,char,pthread\_create,pthread\_t,pthread\_join,double}]

['bcrypt_genblock', 'bcrypt_hash', 'bcrypt_setup', 'openssh_bcrypt']

\end{lstlisting}
\end{minipage}
\caption{\textit{esbmc-wr} function listing output.}
\label{list-func-out}
\end{figure}

\subsection{Running the Model Checker}

Figure~\ref{run-esbmc} shows the \textit{run\_esbmc} function implementation. First, it checks whether all functions will be verified. If so, it lists all function on \textit{c\_file}. The script then runs the model checker, calling a subprocess of ESBMC, indicating which file and function should be checked and which parameters must be used. In the end, it gets the subprocess output and prints it to generate the logs.  

\begin{figure}[ht]
\centering
\begin{minipage}{0.45\textwidth}
\begin{lstlisting}[stepnumber=1,
numbers=left,
firstnumber=1,
numberstyle=\tiny,
extendedchars=true,
breaklines=true,
frame=single,
basicstyle=\footnotesize,
stringstyle=\ttfamily,
showstringspaces=false,
captionpos=b,
breakautoindent=truem
language=python,
numbersep=5pt,
tabsize=2,
morekeywords={assert,assume,int,void,NULL,return,for,malloc,include,if,else,free,sizeof,define,struct,while,typedef,goto,short,char,pthread\_create,pthread\_t,pthread\_join,double}]
def run_esbmc(c_file, cmd_line, dep_list, time, func, witness):
    print("[FILE] ",c_file)

    if not func:
        func_list = ["main"]
		else:
    		func_list = list_functions(c_file)
    		esbmc_args = shlex.split(cmd_line);

    for item in func_list:
        print("[FUNCTION] ",item, "\n") 

        output = subprocess.Popen([ESBMC, c_file] + 
            ([] if not func else [FUNCTION, item]) +
            ([] if not witness else [WITNESS, DIRECTORY + "/" + "graphML_" + item]) +
            esbmc_args + 
            dep_list,
            stdout=subprocess.PIPE,
            stderr=subprocess.PIPE,
            text=True)

        (stdout, stderr) = output.communicate()

        print(stdout)
        print(stderr)



\end{lstlisting}
\end{minipage}
\caption{The \textit{run\_esbmc} function implementation in \textit{esbmc-wr}.}
\label{run-esbmc}
\end{figure}

\subsection{Exporting the results}

\begin{figure}[ht]
	\centering
	\includegraphics[width=8.5cm, frame]{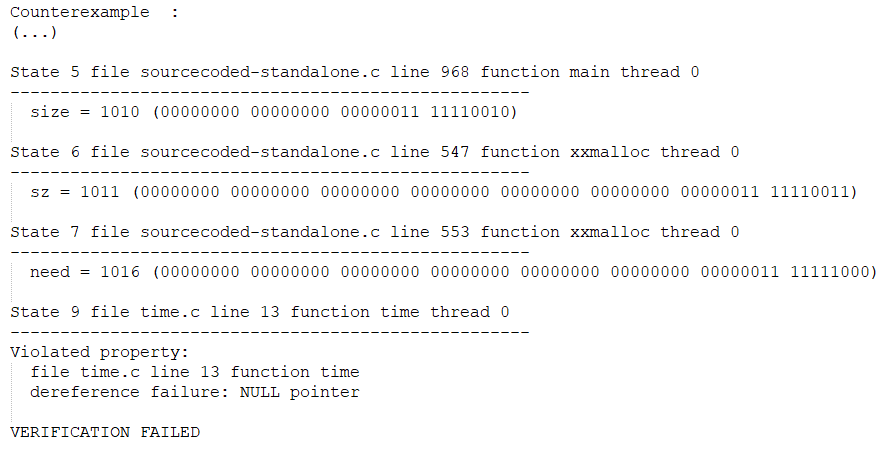}
	\caption{Example of ESBMC counterexample output}
	\label{counterexample}
\end{figure}

After verifying each function, \textit{esbmc-wr} reads the output and writes the relevant pieces of information in a spreadsheet. An example of the model checker output can be found in Figure~\ref{counterexample}.

This counterexample shows a failure found by the model checker. The script will write the following data into the spreadsheet. (1) filename, i.e., \textit{slimguard-standalone.c}~\cite{slim}. (2) The verification status, i.e., \textit{failed}. (3) The function name in which the violation was found, i.e., \textit{time}. (4) The line number in which the function is called, i.e., $13$. (5) The violation type, i.e., \textit{NULL Pointer}.

Using ``-w'' script argument, it is possible to generate GraphML file~\cite{Dirk2015} of each vulnerability found in order to use CPAchecker~\cite{0001S20} to perform a counterexample validation.

\subsection{Illustrative Example}

As an illustrative example, this section describes how to use \textit{esbmc-wr} script. To check an application with several files, e.g., PuTTY \cite{putty}, we have to run \textit{esbmc-wr} on the directory, where the source code is located. The \textit{esbmc-wr} will list all $.c$ files on that directory.

Some applications could have a specific dependency on each file, so it is necessary to list the dependencies before using \textit{esbmc-wr} manually. All dependency paths must be included on a file and provided to \textit{esbmc-wr} using its \textit{-i} parameter. Figure~\ref{depfile} illustrates an example of the dependency file. Each dependency must be included in one line.

\begin{figure}[ht]
\centering
\begin{minipage}{0.45\textwidth}
\begin{lstlisting}[stepnumber=1,
numbers=left,
firstnumber=1,
numberstyle=\tiny,
extendedchars=true,
breaklines=true,
frame=single,
basicstyle=\footnotesize,
stringstyle=\ttfamily,
showstringspaces=false,
captionpos=b,
breakautoindent=truem,
language=C,
numbersep=5pt,
tabsize=2,
morekeywords={assert,assume,int,void,NULL,return,for,malloc,include,if,else,free,sizeof,define,struct,while,typedef,goto,short,char,pthread\_create,pthread\_t,pthread\_join,double}]
 /usr/include/gtk-3.0/
 /usr/include/glib-2.0/
 /usr/lib/x86_64-linux-gnu/glib-2.0/include/
 /usr/include/pango-1.0/
 /usr/include/cairo/
 /usr/include/gdk-pixbuf-2.0/
 /usr/include/atk-1.0/
 ../                
 ../unix/            
 ../charset/          
 /usr/include/harfbuzz/

\end{lstlisting}
\end{minipage}
\caption{An example of dependency file.}
\label{depfile}
\end{figure}

After listing all dependencies, \textit{esbmc-wr} can be run using the parameters described in Section~\ref{sec:SetupConfiguration}. Figure~\ref{script-usage} illustrates an example of how \textit{esbmc-wr} can be used.

\begin{figure}[ht]
\centering
\begin{minipage}{0.45\textwidth}
\begin{lstlisting}[stepnumber=1,
numbers=left,
firstnumber=1,
numberstyle=\tiny,
extendedchars=true,
breaklines=true,
frame=single,
basicstyle=\footnotesize,
stringstyle=\ttfamily,
showstringspaces=false,
captionpos=b,
breakautoindent=truem,
language=C,
numbersep=5pt,
tabsize=2,
morekeywords={assert,assume,int,void,NULL,return,for,malloc,include,if,else,free,sizeof,define,struct,while,typedef,goto,short,char,pthread\_create,pthread\_t,pthread\_join,double}]
    python3 esbmc-wr.py -u 1 -nu -i dep.txt

\end{lstlisting}
\end{minipage}
\caption{An example on how to use the python script esbmc-wr.}
\label{script-usage}
\end{figure}

\section{Experimental Evaluation}
\label{sec:experimental-evaluation}

This section will discuss the experimental evaluation of our proposed approach to verify large software systems using multi-core \textit{k}-induction. In Section~\ref{sec:setup}, we will discuss the setup used in the evaluation. In Sections~\ref{sec:objectives} and~\ref{sec:availability}, we will define our experimental goals and show how one can download and reproduce our experiments. In Sections~\ref{sec:results} and~\ref{sec:threats}, we will assess the obtained results, as well as threats to the validity of our experiments.

\subsection{Experimental Setup}
\label{sec:setup}

All experiments described in this work were carried out on an Intel(R) Xeon(R) CPU E$5$-$2620$ v$4$ @ $2$.$10$GHz with $128$ GB of RAM and Linux OS. We have implemented modifications on top of ESBMC v$6$.$4$.$0$ (publicly available at \url{https://github.com/esbmc/esbmc}) so that it can support the proof by $k$-induction in multi-core architecture that uses incremental verification \url{https://github.com/esbmc/esbmc/pull/236} and the operational models of the libraries used by our benchmarks \url{https://github.com/esbmc/esbmc/pull/280}.
These extensions are currently under review by the ESBMC developers.

All presented execution times are CPU times, {\it i.e.}, only the elapsed periods spent in the allocated CPUs, which was measured with the {\it time} system call~\cite{LinuxManual}. All experimental results reported here were obtained by using our implemented scripts described in Section~\ref{sec:method}.

We selected two applications: PuTTY~\cite{putty} and SlimGuard~\cite{slim}, to evaluate our method to verify large software systems. PuTTY is a free terminal emulator and serial console tool that supports many protocols; some are SSH~\cite{Ylonen2006} and Telnet~\cite{Postel1983}. Note that the PuTTY application and source-code are distributed under the MIT license. It was designed to run on Windows and Unix Platforms~\cite{putty} and is available at \url{https://www.putty.org/}. Slimguard is a secure and efficient memory allocator that focuses on offering a low memory overhead and guaranteeing safe operations as well~\cite{slim}.  We have contacted SlimGuard developers to get its source code to be used in our evaluation.

\subsection{Experimental Objectives}
\label{sec:objectives}

Our experimental evaluation has the following three goals:

\begin{tcolorbox}
\begin{enumerate}

\item[\textbf{EG1}] \textbf{(Correctness)} Evaluate whether our multi-core proof by
	$k$-induction produces correct results faster than the sequential one.

\item[\textbf{EG2}] \textbf{(Soundness)} Evaluate whether our verification approach
	produces results that are confirmed outside our verification engine.

\item[\textbf{EG3}] \textbf{(Scalability)} Evaluate the performance of our verification
	approach when applied to large software systems that are supposed to be
		secure.

\end{enumerate}
\end{tcolorbox}

\subsection{Availability of Data and Tools}
\label{sec:availability}

It is necessary to download both the tools and the dataset used to reproduce our experiments. The tools consist of the developed script, esbmc-wr.py, and ESBMC.\footnote{\url{www.esbmc.org}} The \textit{esbmc-wr.py} source code is available at \url{omitted due to the double-blind review policy.}
Instructions on how to initialize this script can be found on README.md file. 

ESBMC is publicly available under the terms of the Apache License 2.0. Instructions for building ESBMC from the source are given in the file BUILDING (including the description of all dependencies). ESBMC is a joint project with the Federal University of Amazonas (Brazil), University of Southampton (UK), University of Manchester (UK), and University of Stellenbosch (South Africa).

The dataset is available on two different sources. Slimguard can be found at
\url{https://github.com/ssrg-vt/SlimGuard} and PuTTY on
\url{https://git.tartarus.org/simon/putty.git}.

\subsection{Results}
\label{sec:results}

\subsubsection{Multi-core \textit{k}-induction}
\label{sec:threats}

In order to address the correctness of our multi-core proof by $k$-induction, we compared it to ESBMC's conventional sequential proof. We have used one SV-COMP's benchmarks suite, \textit{i.e.}, \textit{AWS C Common}, which is composed of $123$ benchmarks that represent real-world applications at AWS. Table~\ref{table:aws-c-common} summarizes our results.

In Table~\ref{table:aws-c-common}, \textbf{P} denotes the parameter under evaluation, \textbf{A} represents the used approach, \textbf{skind} represents the current sequential approach implemented in ESBMC, \textbf{pkind} denotes the multi-core implementation presented in this work, \textbf{CT} shows the number of benchmarks correctly marked as safe, \textbf{CF} shows the number of benchmarks correctly marked as unsafe, \textbf{IT} denotes the number of benchmarks incorrectly marked as safe, \textbf{IF} denotes the number of benchmarks incorrectly marked as unsafe, \textbf{U} represents the number of benchmarks for which no conclusive results could be obtained, \textit{e.g.}, time/memory limit exhaustion, \textbf{CPU} shows the CPU time spent to evaluate the complete suite in seconds, and finally, \textbf{Wall} shows the wall time spent to evaluate the complete suite in seconds.

\begin{table}[htb]
\caption{Experimental results for AWS C Common suite.}
\label{table:aws-c-common}
\centering
\begin{tabular}{|c|c|c|}
\hline
\textbf{P/A}      & \textbf{skind} & \textbf{pkind} \\ \hline
\textbf{CT}       & 99             & \textbf{108}                 \\ \hline
\textbf{CF}       & 0              & 0                            \\ \hline
\textbf{IT}       & 0              & 0                            \\ \hline
\textbf{IF}       & 0              & 0                            \\ \hline
\textbf{U}        & 24             & \textbf{15}                  \\ \hline
\textbf{CPU (s)} & 1648.71        & \textbf{386.38}              \\ \hline
\textbf{Wall (s)} & 2718.08        & \textbf{1208.42}             \\ \hline
\end{tabular}
\end{table}

As one can note, the approach implemented in this work can compute more correct results and less incorrect results if compared to the ESBMC's conventional sequential proof. Furthermore, in terms of time, our approach was able to verify the complete test suite in less than a third of CPU time and less than half of wall time. 

\begin{tcolorbox}
These results partially answer \textbf{EG1}, since our approach does produce correct results in less time than the sequential approach. However, to completely answer this goal, a more comprehensive comparison must be carried out over a broad set of open-source applications.
\end{tcolorbox}

\subsubsection{Slimguard Verification}

SlimGuard is a secure allocator; therefore, our main objective was to check memory safety properties (e.g., pointer dereference and memory leaks) while performing the verification. To do that, we ran the \textit{esbmc-wr} script with the commands illustrated in Figure~\ref{slim}. As the proposed script lists all ``.c'' files in the current directory, we run it on the same path where SlimGuard was located. The ``-k'' argument enables the ESBMC \textit{k}-induction-parallel parameter, the ``-m'' enables memory leak verification, ``-f'' performs function by function verification and ``-w'' enables the witness output. Note that ESBMC already checks various memory safety properties by default, including object bounds checking and pointer arithmetic. A timeout was set to $300$s to prevent that verification gets stuck. 

\begin{figure}[ht]
\centering
\begin{minipage}{0.45\textwidth}
\begin{lstlisting}[stepnumber=1,
numbers=left,
firstnumber=1,
numberstyle=\tiny,
extendedchars=true,
breaklines=true,
frame=single,
basicstyle=\footnotesize,
stringstyle=\ttfamily,
showstringspaces=false,
captionpos=b,
breakautoindent=truem
language=python,
numbersep=5pt,
tabsize=2,
morekeywords={assert,assume,int,void,NULL,return,for,malloc,include,if,else,free,sizeof,define,struct,while,typedef,goto,short,char,pthread\_create,pthread\_t,pthread\_join,double}]
 esbmc-wr.py -m -k -f -w -t 300

\end{lstlisting}
\end{minipage}
\caption{Command used to verify SlimGuard.}
\label{slim}
\end{figure}

The developers' source code consisted of a single ``.c'' file with $40$ different functions. All functions were verified, but the \textit{main} function led to time out. So, the main function was verified using ESBMC and none vulnerability was found. Therefore, we listed the vulnerabilities found on function-by-function verification.   

Using the verification function-by-function, we have found $27$ security vulnerabilities; most of them were array bounds violation ($6$) and division-by-zero ($6$). Different types of deference failures were found, as well. Table~\ref{slimbefore} shows the number of each vulnerability (N) we have found.

\begin{table}[t]
\begin{tabular}{|l|l|}
	\hline
	\textbf{error}                                            & \textbf{N} \\ \hline
	dereference failure: forgotten memory: dynamic\_1\_array  & 5                       \\ \hline
	division by zero                                          & 5                       \\ \hline
	array bounds violated: array `Class' upper bound          & 4                       \\ \hline
	dereference failure: array bounds violated                & 2                       \\ \hline
	dereference failure: invalid pointer                      & 2                       \\ \hline
	assertion Class{[}index{]}.size != 0                      & 1                       \\ \hline
	assertion node                                            & 1                       \\ \hline
	dereference failure: forgotten memory: dynamic\_11\_array & 1                       \\ \hline
	dereference failure: forgotten memory: dynamic\_15\_array & 1                       \\ \hline
	dereference failure: NULL pointer                         & 1                       \\ \hline
\end{tabular}
\caption{Number of violated properties found in SlimGuard.}
\label{slimbefore}
\end{table}

One of the security vulnerabilities found was a dereference failure related to an invalid pointer on function \textit{next\_entry}. Figure~\ref{entry_fail} illustrates the counterexample that we have found, which was extracted from the output logs. As the \textit{next\_entry} function snippet shown in Figure~\ref{entry_func}, if \textit{cur} pointer was not correctly initialized, it could result in a invalid dereference. Similar issue was found in \textit{print\_list} function shown in Figure \ref{print_fail}. 

\begin{figure}[ht]
\centering
\begin{minipage}{0.45\textwidth}
\begin{lstlisting}[stepnumber=1,
numbers=left,
firstnumber=1,
numberstyle=\tiny,
extendedchars=true,
breaklines=true,
frame=single,
basicstyle=\footnotesize,
stringstyle=\ttfamily,
showstringspaces=false,
captionpos=b,
breakautoindent=truem
language=python,
numbersep=5pt,
tabsize=2,
morekeywords={assert,assume,int,void,NULL,return,for,malloc,include,if,else,free,sizeof,define,struct,while,typedef,goto,short,char,pthread\_create,pthread\_t,pthread\_join,double}]
Counterexample:

State 1 file slimguard-before-fix.c line 925 function next_entry thread 0
-------------------------------------------
Violated property:
	file slimguard-before-fix.c line 925 function next_entry
  dereference failure: invalid pointer

VERIFICATION FAILED

\end{lstlisting}
\end{minipage}
	\caption{Counterexample found on \textit{next\_entry} SlimGuard's function verification.}
\label{entry_fail}
\end{figure}

\begin{figure}[ht]
\centering
\begin{minipage}{0.45\textwidth}
\begin{lstlisting}[stepnumber=1,
numbers=left,
firstnumber=1,
numberstyle=\tiny,
extendedchars=true,
breaklines=true,
frame=single,
basicstyle=\footnotesize,
stringstyle=\ttfamily,
showstringspaces=false,
captionpos=b,
breakautoindent=truem
language=python,
numbersep=5pt,
tabsize=2,
morekeywords={assert,assume,int,void,NULL,return,for,malloc,include,if,else,free,sizeof,define,struct,while,typedef,goto,short,char,pthread\_create,pthread\_t,pthread\_join,double}]
 sll_t* next_entry(sll_t* cur) {
      return cur->next;
 }  

\end{lstlisting}
\end{minipage}
	\caption{Vulnerability found on \textit{next\_entry} SlimGuard function.}
\label{entry_func}
\end{figure}

\begin{figure}[ht]
\centering
\begin{minipage}{0.45\textwidth}
\begin{lstlisting}[stepnumber=1,
numbers=left,
firstnumber=1,
numberstyle=\tiny,
extendedchars=true,
breaklines=true,
frame=single,
basicstyle=\footnotesize,
stringstyle=\ttfamily,
showstringspaces=false,
captionpos=b,
breakautoindent=truem
language=python,
numbersep=5pt,
tabsize=2,
morekeywords={assert,assume,int,void,NULL,return,for,malloc,include,if,else,free,sizeof,define,struct,while,typedef,goto,short,char,pthread\_create,pthread\_t,pthread\_join,double}]
void print_list(sll_t *slist) {
	if (slist == NULL)
		return;

	sll_t * current = slist;

	while (current != NULL) {
		Debug("curr %p\n", (void *)current);
		current = current->next;
	}
} 

\end{lstlisting}
\end{minipage}
	\caption{Vulnerability found on \textit{next\_entry} SlimGuard function.}
\label{print_fail}
\end{figure}

A counterexample validation was performed using existing tools to ensure that vulnerabilities we have found are real ones. The first tool we used was CPAchecker~\cite{0001S20}, which is a comprehensive tool for configurable software verification and contains a module to validate counterexamples~\cite{Dirk2015}, \textit{i.e.}, witness validation, and is used in SV-COMP to validate the results of the competition. Moreover, model checkers implement an option to output the counterexample as a GraphML file~\cite{Dirk2015} and CPAchecker uses such file, the source-code, and a specification file, \textit{i.e.}, a definition of the type of property violation in the program. CPAchecker can validate whether such a counterexample is reachable or not w.r.t. the specification file. Since ESBMC regularly participates in SV-COMP, it already implements this output option and we enabled it when verifying source-files. However, when we applied the witness validation to SlimGuard's counterexamples, CPAchecker failed to prove or falsify them, thereby throwing an error, for two main reasons: ($1$) this approach was implemented to cover benchmarks from SV-COMP~\cite{Dirk2015} and has been used primarily in this scope; and, ($2$) it is not possible to validate properties for counterexamples generated by the verification of specific functions only, and not the complete source-code, which is one of the core ideas in our approach.

The second approach we used to try to validate our counterexamples was eZProofC~\cite{understanding}. It automates the collection and manipulation of counterexamples to generate a new instantiated code (copying the original code and replacing variables assignments using the given values identified by ESBMC) to reproduce the identified error. Therefore,  eZProofC provides the developers with an automated method to confirm the results and, additionally, alleviates the process of analyzing large counterexamples. Thus, we can check whether the faulty behavior is observed in practice. We applied eZProofC to the counterexamples generated, and we were able to reproduce the security errors reported in Table~\ref{slimbefore}. In particular, $3$ out of $22$ instantiated programs lead to a crash, thereby violating the application’s availability principle.

Lastly, we tried to contact the SlimGuard application developers to share our findings and get feedback from them, whether the vulnerabilities we have found are real ones. They did confirm the issues and implemented fixes for them.

\begin{tcolorbox}
Ultimately, these approaches help us answer \textbf{EG2}, since we were able to confirm our results outside our verification engine, both by using an existing counterexample validator and by contacting the application developers.
\end{tcolorbox}

\subsubsection{PuTTY Verification}

PuTTY application code available at \url{https://git.tartarus.org/simon/putty.git} was used to evaluate the script's performance in verifying an extensive application. We verified the most recent tag (0.74). 

In order to run ESBMC with the proposed script, the command illustrated in Figure~\ref{putty0.52} was used. The ``-m'' allows ESBMC to check memory leak, ``-o'' allows it to check overflow, ``-u n'' sets the number \textit{n} of unwinding loops and ``nu'' disables the unwinding assertions. The file (dep.txt) used on ``-i'' parameter contains all PuTTY library dependencies. 

\begin{figure}[ht]
\centering
\begin{minipage}{0.45\textwidth}
\begin{lstlisting}[stepnumber=1,
numbers=left,
firstnumber=1,
numberstyle=\tiny,
extendedchars=true,
breaklines=true,
frame=single,
basicstyle=\footnotesize,
stringstyle=\ttfamily,
showstringspaces=false,
captionpos=b,
breakautoindent=truem
language=python,
numbersep=5pt,
tabsize=2,
morekeywords={assert,assume,int,void,NULL,return,for,malloc,include,if,else,free,sizeof,define,struct,while,typedef,goto,short,char,pthread\_create,pthread\_t,pthread\_join,double}]
 esbmc-wr.py -m -o -u 1 -nu -i dep.txt

\end{lstlisting}
\end{minipage}
\caption{Command used to verify PuTTY}
\label{putty0.52}
\end{figure}

We were able to verify the code function-by-function. Therefore, we verified $2803$ functions on $136$ different `.c' files in $86.11$ minutes. The verification resulted in several violated properties ($1269$), most of them related to deference failure ($969$). Table~\ref{puttytable} shows the top ten property violations found. However, we have not applied the counterexample validators here since PuTTY contains too many source files and those validators are not yet prepared to handle applications with various source files.

\begin{table*}[t]
\centering
\begin{tabular}{|l|c|}
\hline
\textbf{error}                                                                                                                                   & \textbf{\# Of occurrences} \\ \hline
dereference failure: invalid pointer                                                                                                             & 969                        \\ \hline
arithmetic overflow on sub																																																											 & 26                         \\ \hline
dereference failure: NULL pointer                                                                                                                & 25                         \\ \hline
arithmetic overflow on add                                                                                                                       & 23                         \\ \hline
array bounds violated: array 'subkeytypes' upper bound                                                                                           & 18                         \\ \hline
arithmetic overflow on shl                                                                                                                       & 13                         \\ \hline
array bounds violated: array 'subkeytypes' upper bound                                                                                           & 12                         \\ \hline
assertion (len \& 7) == 0                                                                                                                        & 11                          \\ \hline
arithmetic overflow on mul                                                                                                                       & 9                          \\ \hline
assertion ramdom seedstr                                                                                                                         & 8                          \\ \hline
\end{tabular}
\caption{Top ten property violations found in PuTTY application tag 0.74} 
\label{puttytable}
\end{table*}

\begin{tcolorbox}
These results answer \textbf{EG3}, since we could successfully verify an extensive software system within a reasonable time. However, \textbf{EG3} led us to conclude that further work needs to be done to make the counterexample validators useful for developers of open-source applications that contain various source files.
\end{tcolorbox}

\subsection{Threats to Validity}
\label{sec:threats}

\noindent \textit{Benchmark selection}: We report an assessment of our approach to finding vulnerabilities in large software systems over a set of open-source benchmarks. Nevertheless, this set of benchmarks is limited within this paper's scope, and the performance may not generalize to other benchmarks.

\noindent \textit{Performance and correctness of our multi-core proof by $k$-induction}: The implemented multi-core proof by $k$-induction relies on the idea that each step can be computed independently, and such a statement might lead to three times faster verifications, which may not hold on other benchmarks. However, in future work, this approach should be assessed on a broader benchmark set.

\noindent \textit{Counterexample validation}: We have used other studies to validate the obtained counterexamples from our pragmatic approach to verify large software systems. Nonetheless, we rely on the correctness of such tools; thus, in future work, we should assess other approaches to address a broader scope in this matter.

\section{Related Work}
\label{sec:related-work}

The C programming language is widely used to develop critical software, such as operating systems, drivers, and encryption libraries. However, it lacks protection mechanisms, leaving memory and resource management's responsibility in the developer's hands. Any lapse in this regard can result in undefined behavior, which exposes the program to security vulnerabilities. In this way, several studies address the use of automatic tools for checking memory safety properties in C programs~\cite{related-1, balan, related-2, Richardson2020, Rocha2020}.

In~\cite{related-2}, the authors seek to assess state-of-the-art techniques, thus analyzing the performance of different automatic vulnerability detection tools for C programs. For this purpose, a database containing approximately $700$ test cases, representing security-related vulnerabilities was used, previously classified with related memory safety errors. The general results indicate that the verification tools provide adequate support for detecting problems arising from improper use of memory and undefined behaviors, thus attesting to this type of tool's reliability to improve the C programming language codes. However, the authors limited exploring the tools to known test cases without testing a program itself.

Another approach to exploit security vulnerabilities in software is fuzzing~\cite{Bohme2017}. In this matter, the authors in~\cite{Rocha2020} present Map2Check, which is a software verification tool that uses fuzzing, symbolic execution, and inductive invariants to check safety properties in C programs. Furthermore, it uses the LLVM compiler infrastructure to instrument source code, monitors data from the program's execution, and then uses an iterative deepening approach using fuzzers and symbolic execution engines to check such properties \cite{Rocha2020}. However, even though their experimental results show that Map2Check can be useful to verify pointer safety-related properties, it has only been evaluated under SV-Comp benchmarks and not on large software systems.

In~\cite{Richardson2020}, the author describes how to use Capability Hardware Enhanced RISC Instructions (CHERI) to provide memory safety in C/C++ programs. In particular, they show how to overcome memory errors, such as spatial safety violations, because the bounds in memory of an object are ignored. Moreover, the author presents CheriSH, a technique that protects against buffer overflows between the same object fields, enabling complete spatial memory protection in CHERI~\cite{Richardson2020}. Nonetheless, CHERI must be adopted by developers and industries, so that processor architectures can benefit from such mechanism, and CheriSH primarily focuses on pointer safety, not tackling other types of violations, \textit{e.g.}, memory leaks.

Regarding tool properties, the literature shows that BMC has already been applied successfully to discover flaws in real systems and has been extended to support multi-threaded software systems~\cite{understanding, Background}. In~\cite{1st}, the authors investigate SMT-based verification of ANSI-C programs, focusing on embedded software, thereby offering the first SMT-based BMC assessment in industrial applications. Through the results reported by the authors, it is concluded that the ESBMC outperforms the results of the CBMC and SMT-CBMC when considering the verification of embedded software. Other studies have sought to extend or improve existing tools, such as those described in~\cite{pericles} and~\cite{6693074}.

Recently, the authors in~\cite{Cook2020} present the use of model checkers to triage the severity of security bugs at the cloud service provider at the Amazon Web Services (AWS). The authors tackle the severity of bugs discovered/reported in the Xen hypervisor in the case study, used across several AWS servers and services~\cite{Cook2020}. Whenever a bug is reported, engineers should evaluate the bug's potential threat and how quickly they need to be fixed w.r.t. the severity of such. To do so, the authors have applied code transformations in the original source-code and implemented modifications to the C Bounded Model Checker (CBMC)~\cite{Clarke2004} to slice the program under verification and generate a reduced version. The model checker can easily verify the resulting program and the resulting counterexamples can help engineers write security tests to analyze bugs further. However, several abstractions performed in the verification approach might cause the model checker to miss traces and not automatically falsify spurious traces.

\section{Conclusions and Future Work}
\label{sec:conclusion}

This paper presents a BMC approach to verify large software systems considering all application code files using a state-of-the-art bounded model checker. Our proposal resulted in creating a python script that can be used in association with ESBMC to assist in the verification of extensive programs. 

The script checks the files that make up the application one by one, identifying the functions listed in each file and then executes the ESBMC verification, thus seeking to identify vulnerabilities and generating a report summarizing the output. We have also developed a multi-core version of the \textit{k}-induction proof rule to exploit the availability of multi-core machines. 

We evaluated our methods using two different datasets, PuTTY and Slimguard applications. Regarding dataset evaluation, we were able to reproduce known flaws on the PuTTY application and find new ones in Slimguard, confirmed by the tool developer. We have also exploited existing counterexamples validators to confirm the security vulnerabilities we have found in the Slimguard application. Lastly, we have shown that the multi-core \textit{k}-induction proof rule can speed up the verification process by up to three times.

We will improve our verification method for future work in two directions. First, automatically select the parameters to be used by the model checker using machine learning techniques and exploit the clusters' availability to further speed up the verification process. We will also develop a more robust counterexample validator. We want to confirm the counterexample produced by open-source applications that consist of many source files.

\bibliographystyle{unsrt}
\bibliography{citations}

\end{document}